\documentstyle[aps,prb]{revtex}
\begin{document}

\twocolumn[\hsize\textwidth\columnwidth\hsize\csname
@twocolumnfalse\endcsname

\title{First-principles calculations of hot-electron lifetimes in
metals}
\author{I. Campillo$^{1}$, V. M.  Silkin$^{2}$, J. M. Pitarke$^{1,4}$, 
E. V. Chulkov$^{2,4}$, A. Rubio$^{3}$, and P. M. Echenique$^{2,4}$}
\address{$^1$ Materia Kondentsatuaren Fisika Saila, Zientzi Fakultatea, 
Euskal Herriko Unibertsitatea,\\ 644 Posta kutxatila, 48080 Bilbo, Basque 
Country,
Spain\\
$^2$ Materialen Fisika Saila, Kimika Fakultatea, Euskal Herriko
Unibertsitatea,\\
1072 Posta kutxatila, 20080 Donostia, Basque Country, Spain\\
$^3$Departamento de F{\'{\i}}sica Te\'orica, Universidad de Valladolid,
47011 Valladolid, Spain\\
$^4$Donostia International Physics Center (DIPC) and Centro Mixto
CSIC-UPV/EHU,\\
Donostia, Basque Country, Spain}

\date\today

\maketitle

\begin{abstract}
First-principles calculations of the inelastic lifetime of low-energy electrons
in Al, Mg, Be, and Cu are reported. Quasiparticle damping rates are evaluated
from the knowledge of the electron self-energy, which we compute within the
GW approximation of many-body theory. Inelastic lifetimes are then obtained
along various directions of the electron wave vector, with full inclusion of
the band structure of the solid. Average
lifetimes are also reported, as a function of the electron energy. In Al
and Mg, splitting of the band structure over the Fermi level yields
electron lifetimes that are smaller than those of electrons in a free-electron
gas. Larger lifetimes are found in Be, as a result of the characteristic
dip that this material presents in the density of states near the Fermi level.
In Cu, a major contribution from $d$ electrons participating in the
screening of electron-electron interactions yields electron lifetimes which are
well above those of electrons in a free-electron gas with the electron density
equal to that of valence ($4s^1$) electrons.

\end{abstract}
\pacs{PACS numbers: 71.45.Gm, 78.47.+p}
]


\section{Introduction}

Low-energy excited electrons in metals, with energies larger than $\sim
0.5\,{\rm eV}$ above the Fermi energy, experience strong electron-electron (e-e)
scattering processes. Although inelastic lifetimes of these so-called hot
electrons have been investigated for many years on the basis of the
free-electron-gas (FEG) model of the
solid,\cite{QF,Ritchie59,Quinn62,Ashley1,Kleinman,Shelton,Penn0,Tung2,Penn1,Penn2}
time-resolved two-photon photoemission (TR-2PPE) experiments have shown the key
role that band-structure effects may play in the decay
mechanism.\cite{Schmu,Hertel,Petek0,Cao1,Aes1,Knoesel1,Goldm,Cao2,Aes98,Petek3}
First-principles calculations of hot-electron lifetimes that fully include the
band structure of the solid have been reported only very recently for aluminum
and copper.\cite{Igorprl,Ekardt} These calculations\cite{Igorprl} show that
actual lifetimes are the result of a delicate balance between localization,
density of states, screening, and Fermi-surface topology, even in the case of a
free-electron-like metal such as aluminum.

In this paper, we report first-principles calculations of inelastic
lifetimes of excited electrons in a variety of real metals. We start with
free-electron-like trivalent (Al) and divalent (Mg) metals, and then focus on
divalent Be and the role that $d$ electrons play in a noble metal like Cu. First,
we expand the one-electron Bloch states in a plane-wave basis, and solve the
Kohn-Sham equation of density-functional theory (DFT)\cite{Kohn} by invoking
the local-density approximation (LDA) for exchange and correlation (XC). The
electron-ion interaction is described by means of non-local, norm-conserving
ionic pseudopotentials, and we use the one-electron
Bloch states to evaluate the screened Coulomb interaction in the random-phase
approximation (RPA).\cite{Fetter} We finally evaluate the lifetime of an excited
Bloch state from the knowledge of the imaginary part of the
electron self-energy, which we compute within the
GW approximation of many-body theory.\cite{Hedin} Our calculations indicate
that scattering rates may strongly depend, for a given electron energy, on the
direction of the wave vector of the initial state. Also, average lifetimes, as
obtained by averaging over all wave vectors and bands with the same energy,
are found to deviate considerably from those derived for a FEG.

The rest of this paper is organized as follows: Explicit expressions for the
electron decay rate in periodic crystals are derived in section II, within the
GW approximation of many-body theory. Calculated inelastic lifetimes of
hot electrons in Al, Mg, Be, and Cu are presented in section III, and the
conclusions are given in section IV. Atomic units are used throughout, i.e.,
$e^2=\hbar=m_e=1$.

\section{Theory}

Take an inhomogeneous electron system. In the framework of many-body
theory,\cite{Fetter} the damping rate $\tau_i^{-1}$ of an excited electron in the
state $\phi_i({\bf r})$ with energy $E_i$ is obtained from the knowledge of the
imaginary part of the electron self-energy, $\Sigma({\bf r},{\bf r}';E_i)$, as
\begin{equation}\label{eq1}
\tau_i^{-1}=-{2}\int{\rm d}{{\bf r}}\int{\rm d}{{\bf
r}'}\phi_{i}^*({\bf r}){\rm Im}\,\Sigma({\bf r},{\bf r}';E_i)
\phi_{i}({\bf r}').
\end{equation}

In the GW approximation,\cite{Hedin} one considers only 
the first-order term in a
series expansion of the self-energy in terms of the screened Coulomb 
interaction:
\begin{equation}
\Sigma({\bf r},{\bf r}';E_i)={i\over 2\pi}\int dE\,G({\bf r},{\bf
r}';E_i-E)\,W({\bf r},{\bf r}';E),
\end{equation}
where $G({\bf r},{\bf r}';E_i-E)$ represents the one-particle Green function
and $W({\bf r},{\bf r}';E)$ is the time-ordered screened Coulomb
interaction. After replacing the Green function ($G$) by the zero order
approximation ($G^0$), the imaginary part of the self-energy can be evaluated
explicitly:
\begin{equation}\label{eq5}
{\rm Im}\,\Sigma({\bf r},{\bf r}';E_i)=\sum_f\phi_f^*({\bf r}')
{\rm Im}\,W({\bf r},{\bf r}';\omega)\phi_f({\bf r}),
\end{equation}
where $\omega=E_i-E_f$ represents the energy transfer,
the sum is extended over a complete
set of final states $\phi_f({\bf r})$ with energy $E_f$ ($E_F\le E_f\le
E_i$), $E_F$ is the Fermi
energy, and
\begin{eqnarray}\label{eq2}
W({\bf r},{\bf r}';\omega)=&&v({\bf
r}-{\bf r}')+
\int{\rm d}{\bf r}_1\int{\rm d}{\bf r}_2
v({\bf r}-{\bf r}_1)\cr\cr
&&\times\chi({\bf r}_1,{\bf
r}_2;\omega)v({\bf r}_2-{\bf r}').
\end{eqnarray}
Here, $v({\bf r}-{\bf r}')$ represents the bare Coulomb interaction, and
$\chi({\bf r},{\bf r}';\omega)$ is the density-density correlation function of
the solid.

In the framework of time-dependent density-functional theory
(TDDFT),\cite{Runge,Gross3} the density-density correlation function satisfies
the integral equation
\begin{eqnarray}\label{eq3}
\chi({\bf r},{\bf r}';\omega)&&=\chi^0({\bf r},{\bf r}';\omega)+
\int{\rm d}{\bf r}_1\int{\rm d}{\bf r}_2\chi^0({\bf r},{\bf r}_1;\omega)\cr
\cr &&\times
[v({\bf r}_1-{\bf r}_2)+K^{xc}({\bf r}_1,{\bf r}_2;\omega)]\chi({\bf r}_2,
{\bf r}';\omega),
\end{eqnarray}
where $\chi^0({\bf r},{\bf r}';\omega)$ is the density-density correlation
function of noninteracting Kohn-Sham electrons, as described by the solutions of
the time-dependent counterpart of the Kohn-Sham equation. In usual practice,
these amplitudes are approximated by standard LDA wave functions. The kernel
$K^{xc}({\bf r}_1,{\bf r}_2;\omega)$, which accounts for the reduction in the
e-e interaction due to the existence of short-range exchange-correlation (XC)
effects, is obtained from the knowledge of the XC energy
functional. In the RPA, this kernel is taken to be zero.

For periodic crystals, one may introduce the following Fourier expansion for the
screened interaction of Eq. (\ref{eq2}):
\begin{eqnarray}\label{eq8}
W({\bf r},{\bf r}';\omega)=&&{1\over\Omega}\sum_{\bf q}^{BZ}
\sum_{{\bf G},{\bf G}'}{\rm e}^{{\rm i}({\bf q}+{\bf G})\cdot{\bf r}}
{\rm e}^{-{\rm i}({\bf q}+{\bf G}')\cdot{\bf r}'}\cr\cr
&&\times v_{{\bf G}}({{\bf q}})\epsilon_{{\bf G},{\bf G}'}^{-1}({\bf
q},\omega),
\end{eqnarray}
where the first sum is extended over the first Brillouin
zone (BZ), ${\bf G}$ and ${\bf G}'$ are reciprocal lattice
vectors, $\Omega$ is the normalization volume, $v_{{\bf G}}({\bf q})$ represent
the Fourier coefficients of the bare Coulomb interaction, and $\epsilon_{{\bf
G},{\bf G}'}^{-1}({\bf q},\omega)$ are the Fourier coefficients of the inverse
dielectric function,
\begin{equation}\label{eq12}
\epsilon_{{\bf G},{\bf G}'}^{-1}({{\bf q}},\omega)=\delta_{{\bf G},{\bf G}'}+
\chi_{{\bf G},{\bf G}'}({{\bf q}},\omega)\,v_{{\bf G}'}({\bf q}).
\end{equation}
Within RPA,
\begin{equation}\label{eq13}
\epsilon_{{\bf G},{\bf G}'}({\bf q},\omega)=\delta_{{\bf G},{\bf
G}'}-\chi^0_{{\bf G},{\bf G}'}({\bf q},\omega)\,v_{{\bf G}'}({\bf q}),
\end{equation}
where $\chi_{{\bf G},{\bf G}'}^0({\bf q},\omega)$ are the Fourier coefficients
of the density-density correlation function of non-interacting Kohn-Sham
electrons (see, e.g., Ref.\onlinecite{review}).

After introduction of the Fourier representation of Eq. (\ref{eq8}) into
Eq. (\ref{eq5}), and in the limit that the volume of the system $\Omega$
becomes infinite, one finds the following expression for the damping rate of an
electron in the state $\phi_{{\bf k},n_i}({\bf r})$ with energy $E_{{\bf k},n_i}$:
\begin{eqnarray}\label{eq10}
\tau_i^{-1}={1\over \pi^2}\sum_f
\int_{\rm BZ}{{\rm d}{\bf q}}\sum_{{\bf G},{\bf G}'}&&
{B_{if}^*({\bf q}+{\bf G})B_{if}({\bf q}+{\bf G}')\over
\left|{\bf q}+{\bf G}\right|^2}\cr\cr
&&\times{\rm Im}\left[-\epsilon_{{\bf G},{\bf G}'}^{-1}({\bf
q},\omega)\right],
\end{eqnarray}
where $\omega=E_{{\bf k},n_i}-E_{{\bf k}-{\bf q},n_f}$,
and
\begin{equation}\label{eq11}
B_{if}({\bf q+G})=\int{\rm d}{\bf r}\,\phi_{{\bf k},n_i}^{\ast}({\bf r})\,{\rm
e}^{{\rm i}({\bf q+G})\cdot{\bf r}}\,\phi_{{\bf k}-{\bf q},n_f}({\bf r}).
\end{equation}

Couplings of the wave vector ${\bf q}+{\bf G}$ to wave vectors ${\bf q}+{\bf G}'$ with
${\bf G}\neq{\bf G}'$ appear as a consequence of the existence of electron-density
variations in real solids. If these terms, representing the so-called
crystalline local-field effects, are neglected, one can write
\begin{equation}\label{eq17}
\tau_i^{-1}={1\over\pi^2}\sum_f
\int_{\rm BZ}{{\rm d}{\bf q}}\sum_{\bf G}
{\left|B_{if}({\bf q}+{\bf G})\right|^2\over\left|{\bf q}+{\bf
G}\right|^2}
{{\rm Im}\left[\epsilon_{{\bf G},{\bf G}}({\bf q},\omega)\right]\over
|\epsilon_{{\bf G},{\bf G}}({\bf q},\omega)|^2}.
\end{equation}

Within RPA, Eq. (\ref{eq13}) yields
\begin{eqnarray}\label{eq18}
&&{\rm Im}\left[\epsilon_{{\bf G},{\bf G}}({\bf
q},\omega)\right]={2\pi v_{\bf G}({\bf q})\over\Omega}\sum_{\bf k}^{BZ}
\sum_{n,n'}(f_{{\bf k},n}-f_{{\bf k}+{\bf q},n'})\cr\cr
&&\times|\langle\phi_{{\bf
k},n}|e^{-{\rm
i}({\bf q}+{\bf G})\cdot{\bf r}}|\phi_{{\bf k}+{\bf
q},n'}\rangle|^2\delta(\omega -E_{{\bf k}+{\bf q},n'}+
E_{{\bf k},n}).
\end{eqnarray}
Hence, the imaginary part of $\epsilon_{{\bf G},{\bf G}}({\bf q},\omega)$
represents a measure of the number of states available for real
transitions involving a given momentum and energy transfer ${\bf q}+{\bf G}$
and $\omega$, respectively, which is renormalized by the coupling between
initial and final states. The factor
$\left|\epsilon_{{\bf G},{\bf G}}({\bf q},\omega)\right|^{-2}$ in Eq.
(\ref{eq17}) accounts for the screening in the interaction with the probe
electron. Initial and final states of the probe electron enter through the
coefficients $B_{if}({\bf q}+{\bf G})$.

If all one-electron Bloch states entering both the coefficients
$B_{if}({\bf q}+{\bf G})$ and the dielectric function $\epsilon_{{\bf G},{\bf
G}'}({\bf q},\omega)$ were
represented by plane waves, then Eqs. (\ref{eq10}) and (\ref{eq17}) would
exactly coincide with the GW scattering rate of
excited electrons in a FEG, as obtained by Quinn and
Ferrell\cite{QF} and by Ritchie.\cite{Ritchie59} For hot
electrons with energies very near the Fermi level ($E_i\approx E_F$) this result
yields, in the high-density limit ($r_s<<1$),\cite{rs} the well-known formula of
Quinn and Ferrell,\cite{QF} 
\begin{equation}\label{eqQF}
\tau_i^{QF}=263\,r_s^{-5/2}(E_i-E_F)^{-2}\,{\rm eV}^2\,{\rm fs}.
\end{equation}
For a detailed discussion of the range of validity of this approach, see
Ref.\onlinecite{review}.

We note that the decay $\tau_i^{-1}$ of hot electrons in periodic crystals
depends on both the wave vector ${\bf k}$ and the band index $n_i$ of the
initial Bloch state. Nevertheless, we also define $\tau^{-1}(E)$, as the
average of $\tau^{-1}({\bf k},n)$ over all wave vectors and bands lying with
the same energy in the irreducible wedge of the Brillouin zone (IBZ). Decay
rates of hot electrons lying outside the IBZ are considered by simply using the
symmetry property $\tau^{-1}(S{\bf k},n)=\tau^{-1}({\bf k},n)$, where $S$ represents an
operator of the point group of the crystal.

For the evaluation of the polarizability $\chi_{{\bf G},{\bf
G}'}^0({\bf q},\omega)$ and the coefficients $B_{if}({\bf q}+{\bf G})$, Eq.
(\ref{eq11}), we use the self-consistent LDA eigenfunctions of the one-electron
Kohn-Sham hamiltonian of DFT, which we first expand in a plane wave basis,
\begin{equation}\label{eq20}
\phi_{{\bf k},n}({\bf r})={1\over \Omega}\sum_{{\bf G}}u_{{\bf k},n}({\bf G})
e^{{\rm i}({\bf k}+{\bf G})\cdot {\bf r}}.
\end{equation}
The electron-ion interaction is described by means of non-local,
norm-conserving ionic pseudopotentials,\cite{Silkin84,Troullier} and the XC
potential is obtained in the LDA with use of the Perdew-Zunger\cite{Perdew}
parametrization of the XC energy of Ceperley and Alder.\cite{Ceperley}

Well-converged results have been found with the introduction in Eq. (\ref{eq20})
of kinetic-energy cutoffs of $12$, $6$, and $20\,{\rm Ry}$ for Al, Mg, and Be,
respectively. In the case of Cu, all $4s^1$ and $3d^{10}$ Bloch states have
been kept as valence electrons in the pseudopotential generation, and an
energy-cutoff as large as $75\,{\rm Ry}$ has been required, thereby keeping
$\sim 900$ plane waves in the expansion of Eq. (\ref{eq20}). Though
all-electron schemes, such as the full-potential linearized
augmented plane-wave (LAPW) method,\cite{LAPW} are expected to be better suited
for the description of the response of localized $d$ electrons, the plane-wave
pseudopotential approach has already been successfully incorporated in the
description of the dynamical response of copper.\cite{Igorcu}

Samplings over the BZ required for the evaluation of both the dielectric matrix
and the hot-electron decay rate have been performed on Monkhorst-Pack (MP)
meshes:\cite{Monk} $20\times 20\times 20$ for Al,
$24\times 24\times 10$ for Mg, $24\times 24\times 16$ for Be, and
$16\times 16\times 16$ for Cu. For hot-electron energies under study
($E-E_F\sim 0.5-4.0\,{\rm eV}$), the inclusion of up to 40 bands has been
required, and the sum in Eq. (\ref{eq17}) has been extended over $15$ ${\bf
G}$ vectors of the reciprocal lattice, the magnitude of the maximum momentum
transfer ${\bf q}+{\bf G}$ being well over the upper limit of $\sim 2q_F$
($q_F$ is the Fermi momentum). For the evaluation of hot-electron lifetimes
from Eq. (\ref{eq10}), with full inclusion of crystalline local-field
effects, dielectric matrices as large as $40\times 40$ have been considered.

\section{Results and discussion}

\subsection{Aluminum}

Due to the free-electron-like nature of the energy bands of face-centered cubic
(fcc) aluminum [see Fig. 1], a simple metal with no $d$ bands, the impact of the
band structure on the electronic excitations had been presumed for many years to
be small. However, X-ray measurements\cite{Platzman,Platzman2} and careful
first-principles calculations of the dynamical density-response of this
material\cite{Fleszar,Fleszar2} have shown that band-structure effects cannot be
neglected. Full band-structure calculations of the electronic stopping power of
Al for slow ions have shown that the energy loss is
$\sim 7\%$ larger than that of a FEG.\cite{Igorstop} Our present calculations
indicate that actual hot-electron lifetimes in Al are $\sim 35\%$ smaller than
those of electrons in a FEG.

Our {\it ab initio} GW-RPA calculation of the average lifetime $\tau(E)$ of hot
electrons in Al, as obtained from Eq. (\ref{eq10}) with full inclusion of
crystalline local-field effects, is presented in Fig. 2 by solid circles.
The GW-RPA lifetime of hot electrons in a FEG with the electron density equal to
that of valence electrons in Al ($r_s=2.07$) is exhibited in the same
figure, by a solid line. Our calculations indicate that the lifetime of
hot electrons in Al is, within RPA, smaller than that of electrons in
a FEG with $r_s=2.07$ by a factor of $\sim 0.65$. We have performed
band-structure calculations of Eq. (\ref{eq10}) with and without [see also Eq.
(\ref{eq17})] the inclusion of crystalline local-field corrections, and have
found that these corrections are negligible for electron energies under study.
This is an expected result, since Al crystal does not present strong density
gradients nor special electron density directions (bondings).

In order to understand the origin of band-structure effects on hot-electron
lifetimes in Al, we first focus on the role that the band structure plays in
the creation of electron-hole (e-h) pairs. Hence, we evaluate
hot-electron lifetimes from either Eq. (\ref{eq10}) or Eq. (\ref{eq17}) by
replacing the electron initial and final states in
$|B_{if}({\bf q}+{\bf G})|^2$ by plane waves (plane-wave calculation). The result we
obtain with full inclusion of the band structure of the crystal in the
evaluation of
${\rm Im}\,\epsilon_{{\bf G},{\bf G}'}^{-1}({\bf q},\omega)$ is represented in
Fig. 2 by open triangles. Due to splitting of the band structure over the
Fermi level, new channels are opened for e-h pair production, and band-structure
effects tend, therefore, to decrease the lifetime of very-low-energy
electrons by $\sim 5\%$, as in the case of slow ions.\cite{Igorstop}

In the case of moving ions, differences between actual decay rates and
those obtained in a FEG only enter through the so-called energy-loss matrix,
${\rm Im}[-\epsilon_{{\bf G},{\bf G}'}^{-1}({\bf q},\omega)]$. However,
hot-electron decay rates are also sensitive to the actual initial and final
states entering the coefficients $B_{if}({\bf q}+{\bf G})$.
Differences between our full (solid circles) and plane-wave (open triangles)
calculations come from this sensitivity of hot-electron initial and final states
on the band structure of the crystal, showing that the splitting of the band
structure over the Fermi level now plays a major role in lowering the
hot-electron lifetime.

Scaled lifetimes, $\tau(E)\times(E-E_F)^2$, of hot electrons in Al, as obtained
from our full band-structure calculation (solid circles) and from the FEG model
with
$r_s=2.07$ (solid line), are represented in the inset of Fig. 2. In the limit
$E\to E_F$, the available phase space for real transitions is simply $E-E_F$,
which yields the $(E-E_F)^{-2}$ quadratic scaling of very-low-electron
energies in a FEG, as predicted by Eq. (\ref{eqQF}) (dashed line).\cite{note1}
However, as the energy increases momentum and energy conservation prevents the
available phase space from being as large as $E-E_F$, and the lifetime of
electrons in a FEG departures, therefore, from the $(E-E_F)^{-2}$ scaling. For
energies under study, band-structure effects in Al are found to be nearly
energy-independent; hence, our calculated lifetimes are found to approximately
scale as in the case of electrons in a FEG, and they slightly departure,
therefore, from the $(E-E_F)^{-2}$ scaling. The agreement at $E-E_F\sim
3\,{\rm eV}$ between actual lifetimes and those predicted by Eq. (\ref{eqQF}) is
simply due to the nearly thorough compensation, at these energies, between the
departure of this formula from full free-electron-gas RPA calculations and
band-structure effects. 

Although the energy bands of Al show an overall similarity with the fcc
free-electron band structure, in the vicinity of the W-point there are large
differences between the two cases [see Fig. 1]. At this point, the free-electron
parabola opens an energy gap around the Fermi level and splits along the $WX$
direction into two bands ($Z_1$ and
$Z_4$) with energies over the Fermi level. We have
calculated lifetimes of hot electrons in these bands, with the wave
vector along the $WX$ direction. The results of these calculations are
exhibited in Fig. 2, as a function of energy, by dotted ($Z_1$) and
long-dashed lines ($Z_4$). Although hot electrons in the $Z_1$ band have one more
channel available to decay along the $WX$ direction, the band gap at the
$W$-point around the Fermi level results in hot electrons living longer on the
$Z_1$ than on the $Z_4$ band. When the hot-electron energy is well above the
Fermi level ($E-E_F>4\,{\rm eV}$), both $Z_1$ and $Z_4$ lifetimes nearly
coincide with the average values represented by solid circles. We have
evaluated hot-electron lifetimes along other directions of the wave vector, and
have found that differences between these results and average lifetimes are not
larger than those obtained along the $WX$ direction. This is in
disagreement with the calculations reported by Sch\"one {\it et
al}.\cite{Ekardt} In particular, the bending of the hot-electron lifetime
along the WL direction at $\sim 1\,{\rm eV}$ reported in
Ref.\onlinecite{Ekardt} is not present in our calculations. The origin of this
discrepancy is the crossing near the Fermi level between bands $Q_+$ and $Q_-$
along the WL direction reported in Ref.\onlinecite{Ekardt}. This
crossing is absent in the present [see Fig. 1] and
previous\cite{Silkin84,Singhal,Segall2,Papa} self-consistent band-structure
calculations, which all show that at the $W$-point of the Al band structure the
level $W_2'$ is below $W_1$.    
 
\subsection{Magnesium}

In Fig. 3 we show the band structure of hexagonal closed-packed (hcp)
magnesium. There is a close resemblance for energies $E<E_F$ between this band
structure and that of free electrons, though the free-electron parabola now
splits near the Fermi level along certain symmetry directions. As a
result, the energy-loss function, ${\rm
Im}[-\epsilon_{{\bf G},{\bf G}'}^{-1}({\bf q},\omega)]$, of this material is approximately
well described within a free-electron model, and band-structure effects on
hot-electron lifetimes enter mainly, as in the case of Al, through the
sensitivity of hot-electron initial and final states on the band structure of
the crystal.

Our {\it ab initio} calculation of the average lifetime $\tau(E)$ of hot
electrons in Mg, as obtained from Eq. (\ref{eq10}) with full inclusion of
crystalline local-field effects, is presented in Fig. 4 by solid circles,
together with the lifetime of hot electrons in a FEG with the electron
density equal to that of valence electrons in Mg ($r_s=2.66$). As in the
case of Al, we have found that local-field corrections are negligible for
electron energies under study.

Scaled lifetimes of hot electrons in Mg, as obtained from our full
band-structure calculations (solid circles) and from the FEG model with
$r_s=2.66$ (solid line), are represented in the inset of Fig. 4. We note
that actual lifetimes in this material scale with energy approximately as
in the case of electrons in a FEG, thereby slightly deviating from the
$(E-E_F)^{-2}$ scaling predicted by Eq. (\ref{eqQF}) (dashed line). Because of
splitting of the band structure over the Fermi level new decay channels are
opened, not present in the case of a FEG, and band-structure effects
tend, therefore, to decrease the lifetime of hot electrons in Mg by a factor
of $\sim 0.75$. Since the splitting of the band structure of Mg is not as
pronounced as that of Al, the departure of actual lifetimes from those of
electrons in a FEG is found to be smaller in Mg than in Al.

\subsection{Beryllium}

Among the so-called simple metals, with no $d$ bands, beryllium presents
distinctive features in that its band structure 
[see Fig. 5] exhibits the largest
departure from free-electron behaviour. Both Be and Mg have hcp crystal
structure and two conduction electrons per atom. Nevertheless, the electronic
structure of Be is qualitatively different from that of Mg, the Be band gaps at
points $\Gamma$, $H$, and $L$ being much larger than those in Mg. Also, the Be
band gaps are located on both sides of the Fermi level, and the density of states
(DOS) of this material falls to a sharp minimum near the Fermi level.

Our band-structure calculation of the average lifetime $\tau(E)$ of hot
electrons in Be is shown in Fig. 6 by solid circles, as obtained from Eq.
(\ref{eq10}). The lifetime of hot electrons in a FEG with the electron density
equal to that of valence electrons in Be ($r_s=1.87$) is represented in the
same figure by a solid line. In the inset, the corresponding scaled
calculations are plotted, together with the results obtained from Eq.
(\ref{eqQF}) (dashed line). It can be seen that large deviations from the FEG
calculation occur for electron energies near the Fermi level ($E-E_F<3\,{\rm
eV}$), especially at
$E-E_F\sim 1.4$ and $1.8\,{\rm eV}$ where the presence of band gaps at points
$\Gamma$ and $L$ plays a key role. We note that the deep departure from
free-electron behaviour of the beryllium DOS near the Fermi level tends to
increase the inelastic lifetime of all excited Bloch states. Furthermore, actual
lifetimes strongly deviate from the
$\sim(E-E_F)^{-2}$ scaling predicted within Fermi-liquid theory. This deviation
comes from the contribution to the average lifetime due to Bloch states near the
points $\Gamma$ and $L$ with energies close to the energy gap.

At the $\Gamma$-point, the free-electron parabola opens a wide energy gap
around the Fermi level and splits along the $\Gamma K$ and $\Gamma M$
directions into two bands, $T_2/T_4$ and $\Sigma_1/\Sigma_3$, respectively. The
results of our calculated lifetimes of hot electrons in bands $T_2$ and
$T_4$, with the wave vector along the $\Gamma K$ direction, are plotted in Fig.
7 by short-dashed and long-dashed lines, respectively. For comparison, the
average lifetime of hot electrons in real beryllium and in a FEG with
$r_s=1.87$ are also plotted in this figure by solid circles and by a solid line,
respectively. At very-low electron energies ($E-E_F<1\,{\rm eV}$), interband
transitions yield lifetimes of hot electrons in the $T_2$ band that are below
those of electrons in a FEG, as in the case of Al and Mg. However, at higher
energies the coupling with lower lying flat bands becomes small, and
lifetimes of electrons in this ($T_2$) band are found to be above the FEG
prediction. Lifetimes of hot electrons in the $T_4$ band are also, at
very-low electron energies ($E-E_F<1.4\,{\rm eV}$), below those of
electrons in a FEG. Nevertheless, at energies of $\sim 1.4\,{\rm eV}$ the
presence of the band gap at $\Gamma$ yields very long lifetimes, especially at
the level $\Gamma_4^-$.

We have calculated hot-electron lifetimes in bands
$\Sigma_1$ and $\Sigma_3$, with the wave vector along the $\Gamma M$
direction, and have found results that are similar to those obtained for
electrons in bands $T_2$ and $T_4$. Calculations of the lifetime of hot
electrons in the $\Delta_2$ band, with the wave vector along the $\Gamma
A$ direction, are also shown in Fig. 7 (dotted line). Though this is not a flat
band, the presence of the gap at the $\Gamma$-point near the Fermi level results
in hot electrons close to the level $\Gamma_4^-$ living much longer
than in the case of a FEG. The combined contribution from hot electrons in bands
$T_4$,
$\Sigma_1$, and
$\Delta_2$ is the origin of the enhanced average lifetime (solid circles) at
$E-E_F\sim 1.4\,{\rm eV}$, which corresponds to the energy of the
level $\Gamma_4^-$ at the $\Gamma$-point.

A wide band gap is also opened at the $L$-point, which originates the
enhanced average lifetime at $E-E_F\sim 1.8\,{\rm eV}$. Hence, we have
plotted in Fig. 8 the results of our calculated lifetimes of hot electrons
in the $S_1$ band, with the wave vector along the $LH$ direction
(dotted line), together with the average lifetime of hot electrons in
real Be (solid circles) and in a FEG with $r_s=1.87$ (solid line). As in the
case of hot electrons in bands $T_4$ and $\Sigma_1$, the presence of the band
gap at the $L$-point yields a very long average lifetime, but now at $E-E_F\sim
1.8\,{\rm eV}$. A similar behaviour is obtained near the $L$-point for electrons
in the $R_1R_3$ band along the $LA$ direction, and both bands,
$S_1$ and $R_1R_3$, contribute to the enhanced average lifetime at $E-E_F\sim
1.8\,{\rm eV}$. In Fig. 8 we have also represented calculations of
the lifetime of hot electrons in the $S_1'$ band along the $HA$ direction 
(dashed line). At low energies ($E-E_F<2\,{\rm eV}$), the presence of the gap at
the $H$-point leads to hot-electron lifetimes along the $HA$ direction that are
longer than those of electrons in a FEG, but departure from free-electron
behaviour at the $H$-point is not as pronounced as at the $\Gamma$ or $L$ points.
At higher energies, the $S_1'$ band shows great similarity with the
corresponding hcp free-electron band, and lifetimes nearly coincide,
therefore, with those obtained within the FEG model.      

\subsection{Copper}

Copper, the most widely studied metal by TR-2PPE, is a noble metal with
entirely filled $3d$-like bands. In Fig. 9 we show the energy bands of this
fcc crystal. We see a profound difference between the band structure of Cu
and that of free electrons. Slightly below the Fermi level, at $E-E_F\sim
2\,{\rm eV}$, we have $d$ bands capable of holding 10 electrons per atom,
the one remaining electron being in a free-electron-like band below and
above the $d$ bands. Hence, a combined description of both delocalized
$4s^1$ and localized $3d^{10}$ electrons is needed to address the actual
electronic response of this material. The results presented below have been
found by keeping all $4s^1$ and $3d^{10}$ Bloch states as valence electrons
in the pseudopotential generation.

Band-structure GW-RPA calculations of the average lifetime $\tau(E)$ of hot
electrons in Cu are exhibited in Fig. 10 by solid circles, as obtained from
Eq. (\ref{eq10}) with full inclusion of crystalline local-field effects. The
lifetime of hot electrons in a FEG with the electron density equal to that
of $4s^1$ electrons in Cu ($r_s=2.67$) is represented by a solid line. These
calculations indicate that the lifetime of hot electrons in Cu is, within
RPA, larger than that of electrons in a FEG with $r_s=2.67$, this
enhancement varying from a factor of $\sim 2.5$ near the Fermi level
($E-E_F=1.0\,{\rm eV}$) to a factor of $\sim 1.5$ for $E-E_F=3.5\,{\rm eV}$. In
order to investigate the role that localized $d$ bands play in the decay
mechanism of hot electrons, we have also used an {\it ab initio}
pseudopotential with the $3d$ shell assigned to the core. The result of this
calculation, displayed in Fig. 10 by a dotted line, shows that it nearly
coincides with the FEG calculation; thus, $d$-band states play a key role in the
hot-electron decay.

We have performed band-structure calculations of Eq. (\ref{eq10}) with and
without [see also Eq. (\ref{eq17})] the inclusion of crystalline local-field
corrections, and have found that these corrections are negligible for
$E-E_F>1.5\,{\rm eV}$, while for energies very near the Fermi level neglecting
these corrections results in an overestimation of the lifetime of less than
$5\%$. Therefore, differences between our full band-structure calculations
(solid circles) and FEG calculations (solid line) come from the actual DOS
available for real excitations, localization, additional screening, and
Fermi-surface topology.

First of all, we focus on the role that both DOS and coupling
between Bloch states participating in the creation of e-h pairs, i.e.,
localization [see Eq. (\ref{eq18})] play in the hot-electron decay mechanism.
Hence, we neglect crystalline local-field effects and present the result of
evaluating hot-electron lifetimes from Eq. (\ref{eq17}) by replacing initial and
final states in $|B_{if}({\bf q}+{\bf G})|^2$ by plane waves and the dielectric
function in
$\left|\epsilon_{{\bf G},{\bf G}}({\bf q},\omega)\right|^{-2}$ by that of a FEG with
$r_s=2.67$. If we further replaced ${\rm
Im}\left[\epsilon_{{\bf G},{\bf G}}({\bf q},\omega)\right]$ by that of a FEG, then we would
obtain the FEG calculation represented by a solid line. The impact of the
actual DOS below the Fermi level may be described by simply replacing the
one-electron Bloch states in Eq. (\ref{eq18}) by plane waves but keeping the
actual number of states available for real excitations. The result of this
calculation\cite{note2} is represented in Fig. 10 by a dashed line. This result
is very close to that reported by Ogawa {\it et al},\cite{Petek0} though these
authors approximated the FEG dielectric function in
$\left|\epsilon_{{\bf G},{\bf G}}({\bf q},\omega)\right|^{-2}$ within the static
Thomas-Fermi model.

It is clear from Fig. 10 that the actual DOS available for real
transitions yields lifetimes that are shorter than those obtained in a FEG
model, especially for $E-E_F>2\,{\rm eV}$ due to opening of the $d$-band
scattering channel dominating the DOS with energies $\sim 2\,{\rm eV}$. However,
if one takes into account, within a full description of the band structure of
the crystal in the evaluation of
${\rm Im}\left[\epsilon_{{\bf G},{\bf G}}({\bf q},\omega)\right]$, the actual
coupling between initial and final states avaliable for real transitions, then
one obtains hot-electron lifetimes which lie, at very-low electron energies
($E-E_F<2.5\,{\rm eV}$) just above the FEG curve [see open circles in Fig. 1 of
Ref.\onlinecite{Igorprl}]. This enhancement of the lifetime, even at energies
below the opening of the
$d$-band scattering channel, is due to the fact that states just below
the Fermi level have a small but significant $d$ component, thus
being more localized than pure $sp$ states.

The combined effect of DOS and localization, which enters through the
imaginary part of the dielectric matrix ${\rm
Im}\left[\epsilon_{{\bf G},{\bf G}'}({\bf q},\omega)\right]$, increases the lifetime of hot
electrons with energies $E-E_F<2.5\,{\rm eV}$ [see open circles in Fig. 1 of
Ref.\onlinecite{Igorprl}]. As for the departure of hot-electron initial and
final states from free-electron behaviour, entering through the coefficients
$B_{if}({\bf q}+{\bf G})$, we have found that it yields hot-electron lifetimes
that are strongly directional dependent, Fermi-surface shape effects tending to
decrease the average inelastic lifetime of very-low-energy electrons
($E-E_F<2.5\,{\rm eV}$) [see Ref.\onlinecite{Igorprl}]. Furthermore, the
combined effect of DOS and localization, on the one hand, and Fermi-surface
shape effects, on the other hand, nearly compensate. Consequently, large
differences between hot-electron lifetimes in real Cu and in a FEG with
$r_s=2.67$ are mainly due to a major contribution from $d$ electrons
participating in the screening of electron-electron interactions, which is
accounted through the factor $\left|\epsilon_{{\bf G},{\bf G}}({\bf
q},\omega)\right|^{-2}$ in Eq. (\ref{eq17}).

The Fermi surface of Cu is greatly flattened in certain regions, showing
a pronounced neck in the direction $\Gamma L$. Thus, the isotropy of
hot-electron lifetimes in a FEG disappears in this material. While flattening
of the Fermi surface along the $\Gamma K$ direction is found to decrease the
hot-electron lifetime by a factor that varies from $\sim 15\%$ near the Fermi
level ($E-E_F=1\,{\rm eV}$) to $\sim 5\%$ for $E-E_F=3.5\,{\rm eV}$ [see also
Ref.\onlinecite{Adler}], the lifetime of hot electrons with the wave vector along
the necks of the Fermi surface, in the $\Gamma L$ direction, is found to be much
longer than the average lifetime. We have calculated hot-electron lifetimes in
the
$\Lambda_1$ band, with the wave vector along the $\Gamma L$ direction, and have
found the lifetime of hot electrons at the $L_1$ level with $E-E_F=4.2\,{\rm
eV}$ to be longer than the average lifetime at this energy by $\sim 80\%$.

A comparison between our calculated hot-electron lifetimes in Cu and those
determined from most recent TR-2PPE experiments was presented in
Ref.\onlinecite{Igorprl}. At $E-E_F<2\,{\rm eV}$, our calculations are close to
lifetimes recently measured by Knoesel {\it et al} in the very-low energy
range.\cite{Knoesel1}  At larger electron energies, good agreement between our
band-structure calculations and experiment is obtained for Cu(110),\cite{Petek0}
the only surface with no band gap in the ${\bf k}_\parallel=0$ direction.

\section{Conclusions}

We have presented full GW-RPA band-structure calculations of the inelastic
lifetime of hot electrons in Al, Mg, Be, and Cu, and have demonstrated that decay
rates of low-energy excited electrons strongly depend on the details of the
electronic band structure. Though the dependence of hot-electron lifetimes in Al
and Mg on the direction of the wave vector has been found not to be large, in
the case of Be and Cu hot-electron lifetimes have been found to be strongly
directional dependent. Furthermore, very long lifetimes at certain points of
the BZ in Be yield average lifetimes in this material which strongly deviate from
the $\sim(E-E_F)^{-2}$ scaling predicted within Fermi-liquid theory.  

As far as band-structure effects on hot-electron energies and wave functions
are concerned, we have found that both splitting of the band structure and the
presence of band gaps over the Fermi level play an important role in the
e-e decay mechanism. In Al and Mg, splitting of the band
structure is found to yield electron lifetimes that are
smaller than those of electrons in a FEG. On the other hand, large deviations
of the band structure of Be along certain symmetry directions from the
free-electron model near the Fermi level result in a strong directional
dependence of hot-electron lifetimes in this material.

As for the presence of band-structure effects on the creation of e-h pairs, there
are contributions from the actual DOS available for real transitions,
from localization, i.e., the actual coupling between electron and hole states,
and from screening. The combined effect of DOS and localization is found not to
be large, even in the case of a noble metal like Cu with $d$ bands. However,
large differences between hot-electron lifetimes in Cu and in a FEG with the
electron density equal to that of valence ($4s^1$) electrons are found to be due
to a major contribution from $d$ electrons participating in the screening of e-e
interactions.

Crystalline local-field corrections in these materials have been found to be
small for hot-electron energies under study.

\section{Acknowledgments}

We would like to thank A. G. Eguiluz for stimulating discussions. We
also acknowledge partial support by the University of the Basque Country, the
Basque Hezkuntza, Unibertistate eta Ikerketa Saila, and the Spanish Ministerio de
Educaci\'on y Cultura.

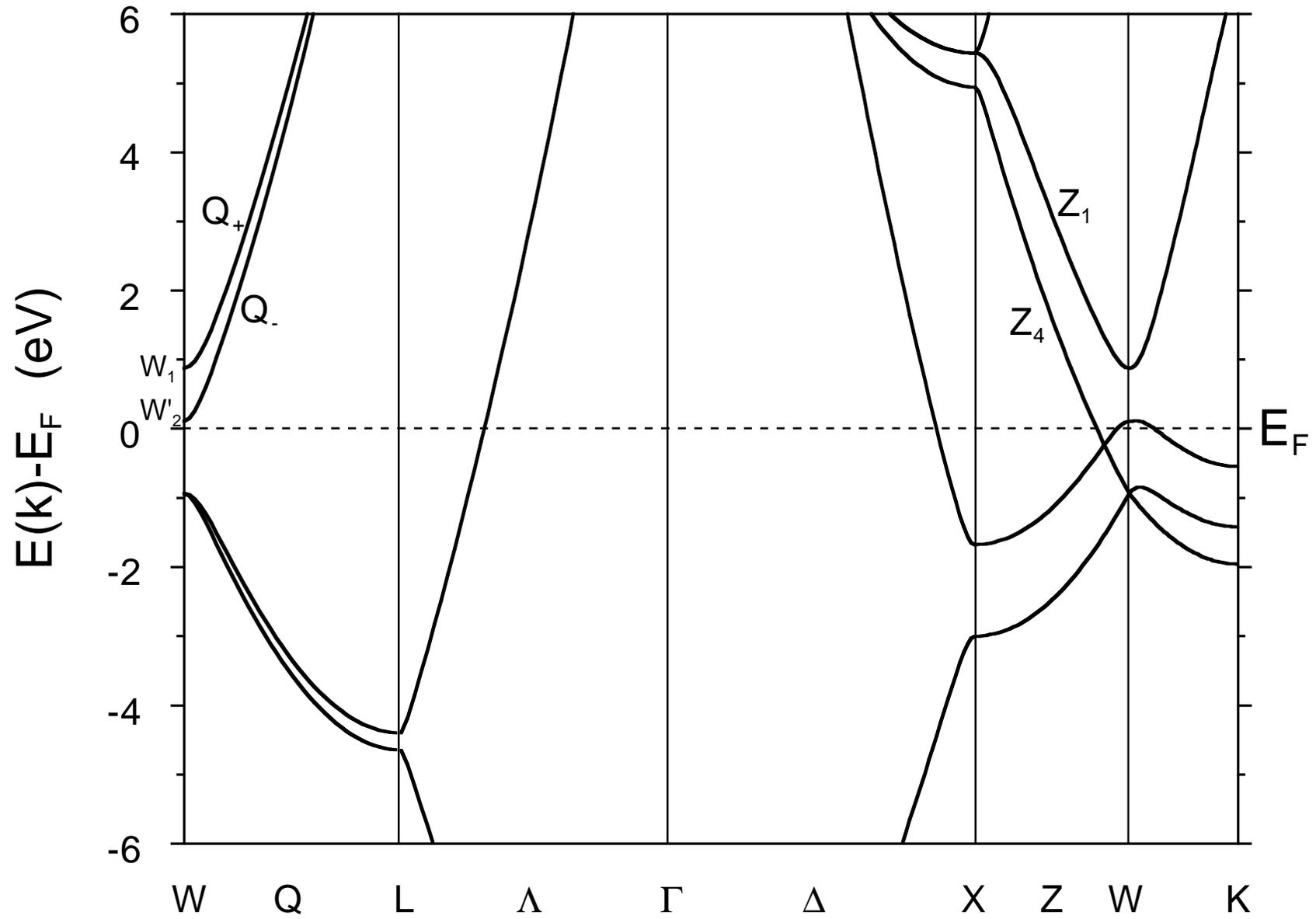
\begin{figure}
\caption[]{Calculated band structure of Al along certain symmetry directions.}
\end{figure}

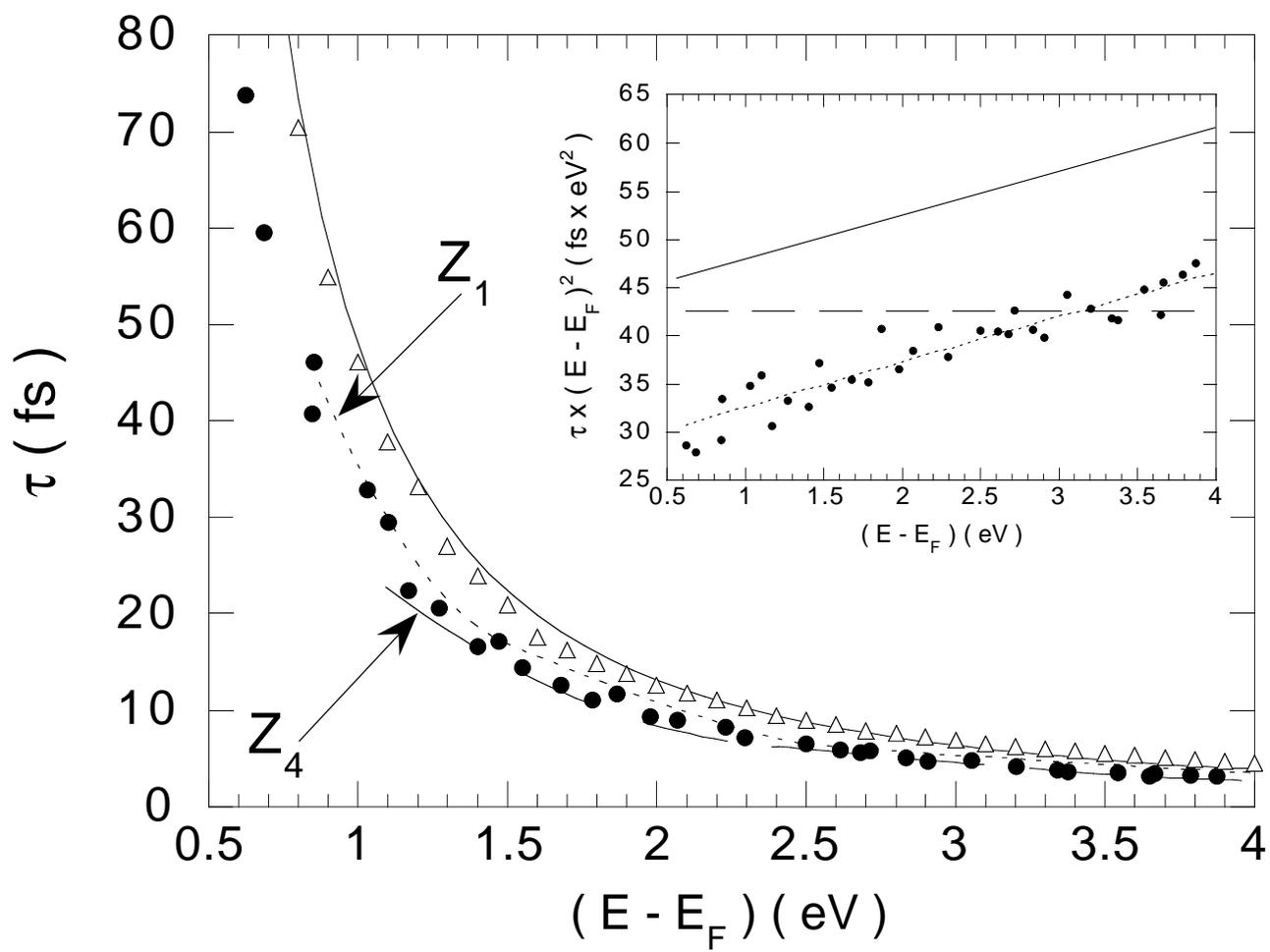
\begin{figure}
\caption[]{Hot-electron lifetimes in Al. Solid circles represent 
our full {\it ab
initio} calculation of $\tau(E)$, as obtained after averaging $\tau({\bf
k},n_i)$ of either Eq. (\ref{eq10}) or Eq. (\ref{eq17}) over wave vectors and
over the band structure for each ${\bf k}$. The solid line represents the
lifetime of hot electrons in a FEG with $r_s=2.07$, as obtained within the full
GW-RPA. Open triangles represent the result obtained from Eq.  (\ref{eq17})
by  replacing
hot-electron initial and final
states in $\left|B_{if}({\bf
q}+{\bf G})\right|^2$ by plane waves, but with full inclusion of the band
structure in the evaluation of ${\rm Im}\left[-\epsilon_{{\bf
G},{\bf G}}^{-1}({\bf q},\omega)\right]$. Dotted and long-dashed lines represent
the lifetime of hot-electrons in bands $Z_1$ and $Z_4$, respectively, with the
wave vector along the $WX$ direction. The inset exhibits scaled lifetimes of
hot electrons in Al. Solid circles and the solid line represent band-structure
and FEG calculations, respectively, both within GW-RPA. The dashed line
represents the prediction of Eq. (\ref{eqQF}).}
\end{figure}

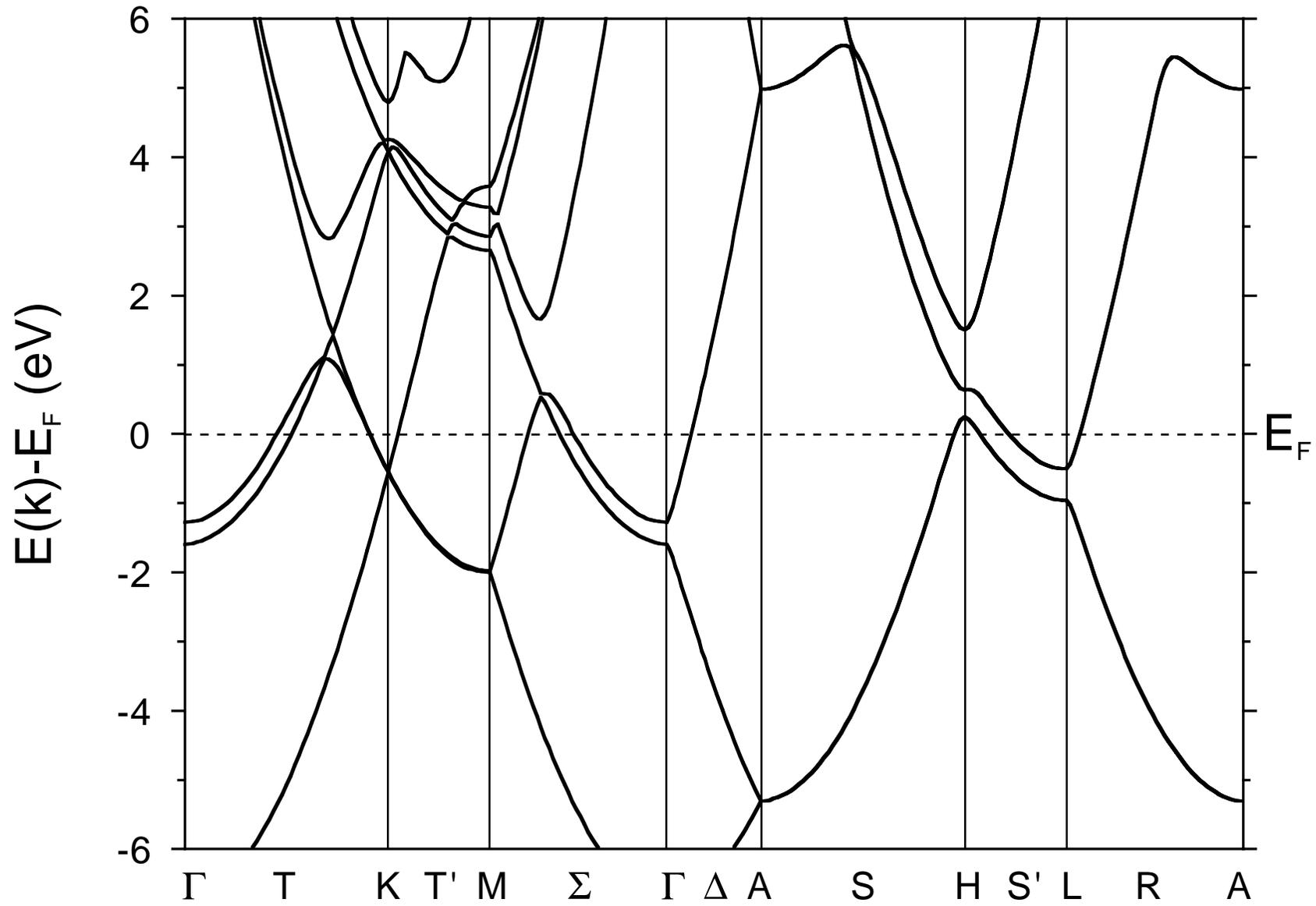
\begin{figure}
\caption[]{Calculated band structure of Mg along certain symmetry directions.}
\end{figure}

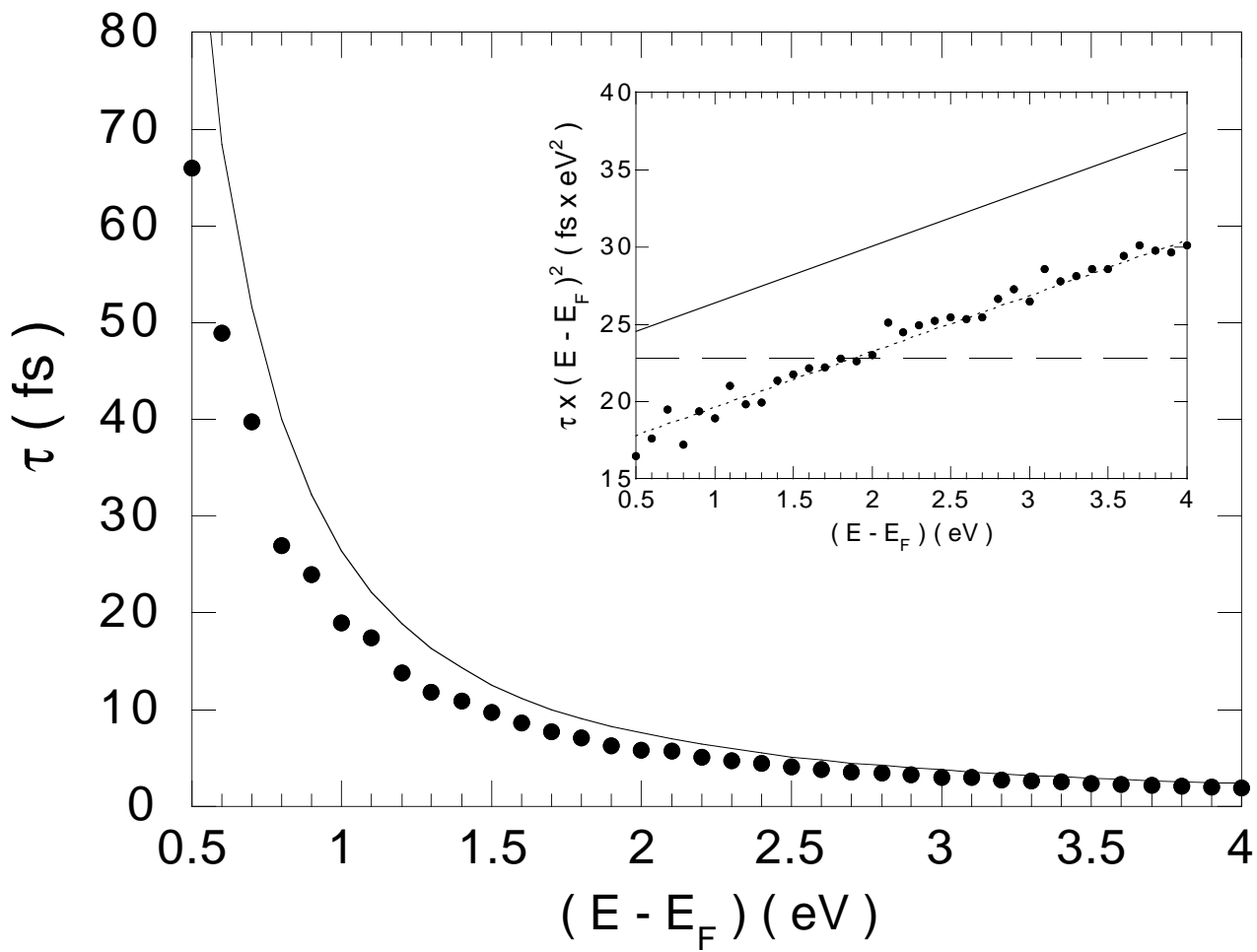
\begin{figure}
\caption[]{Hot-electron lifetimes in Mg. As in Fig. 2 with $r_s=2.66$.}
\end{figure}

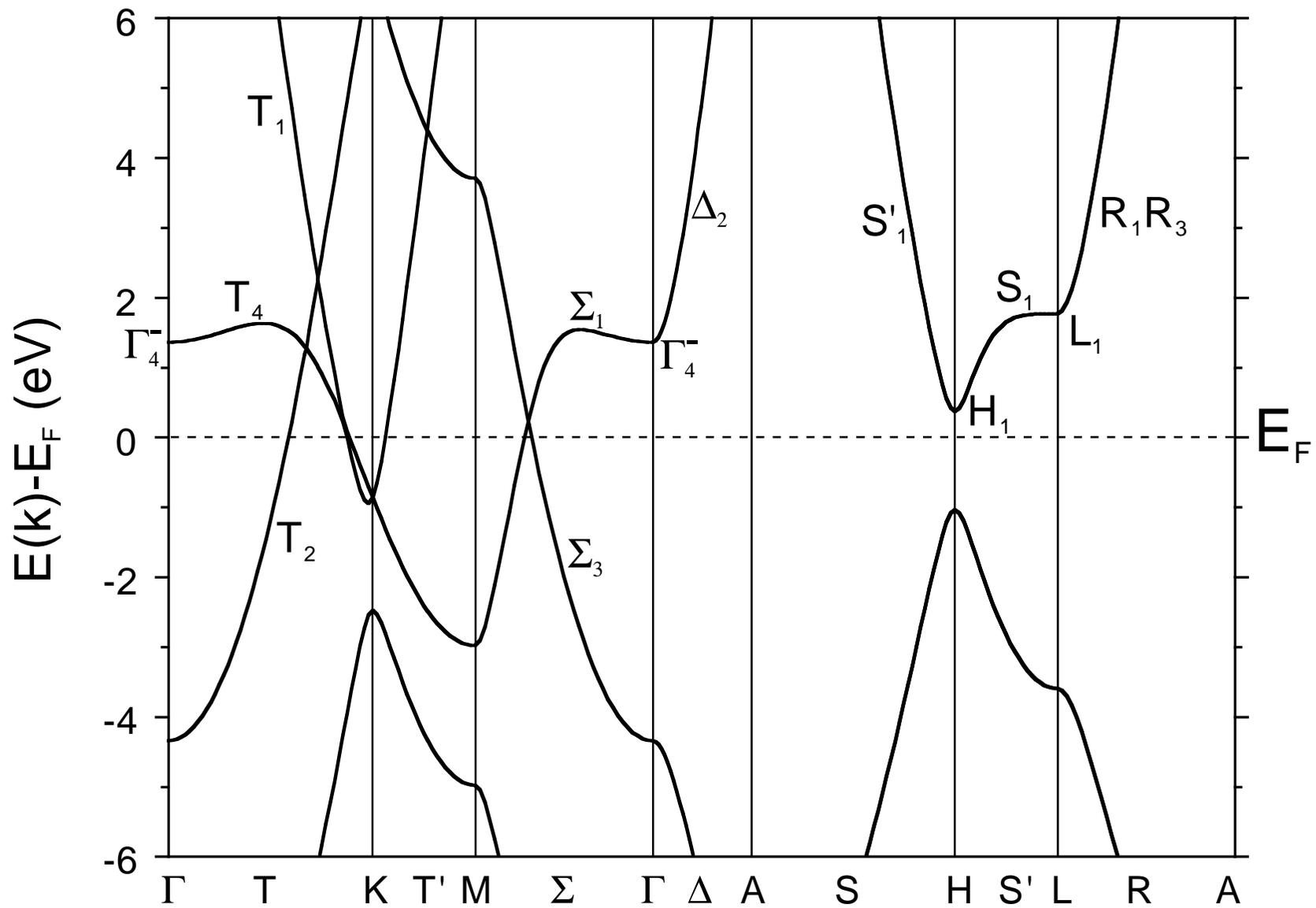
\begin{figure}
\caption[]{Calculated band structure of Be along certain symmetry directions.}
\end{figure}

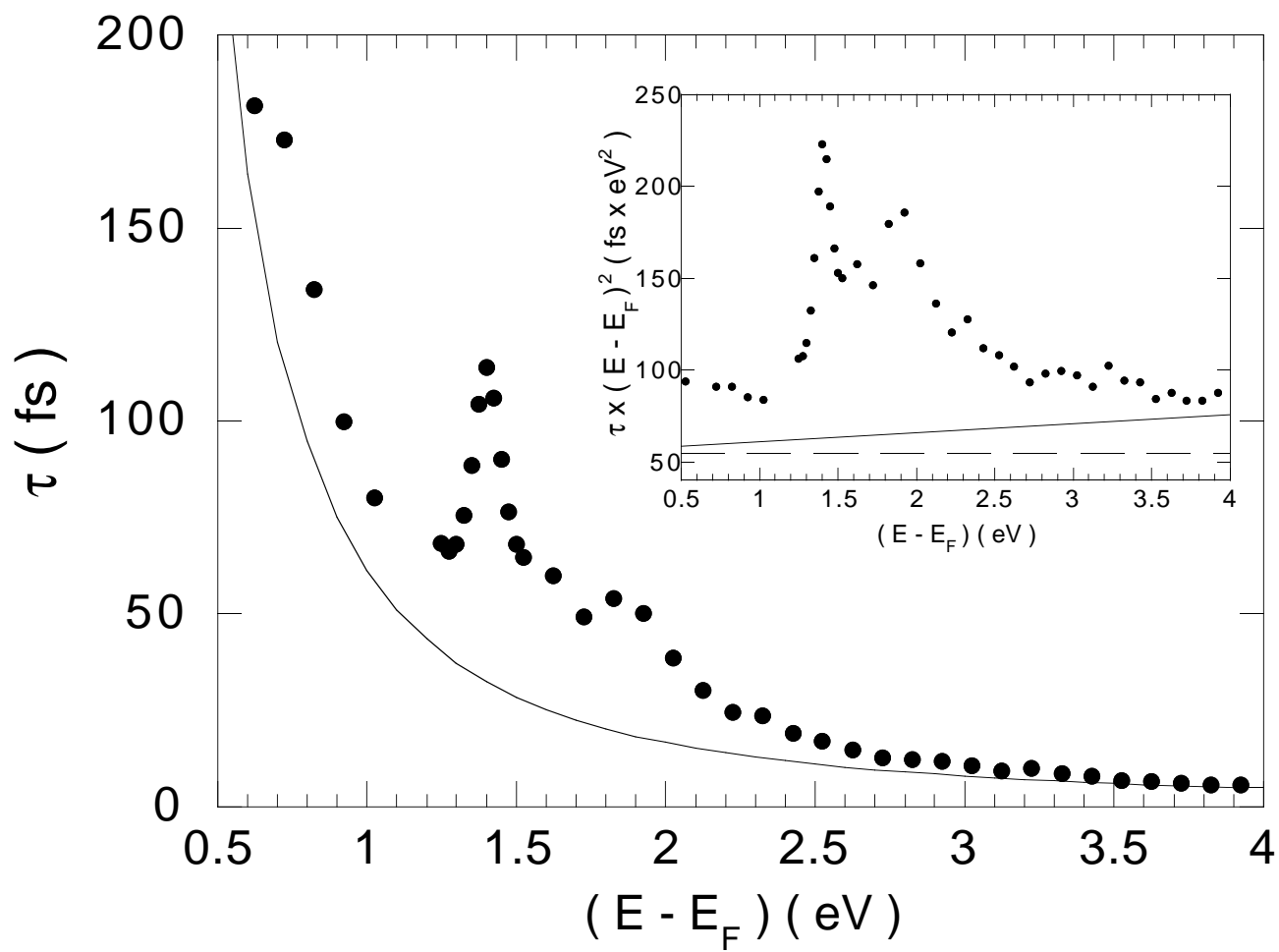
\begin{figure}
\caption[]{Hot-electron lifetimes in Be. As in Fig. 2 with $r_s=1.87$.}
\end{figure}

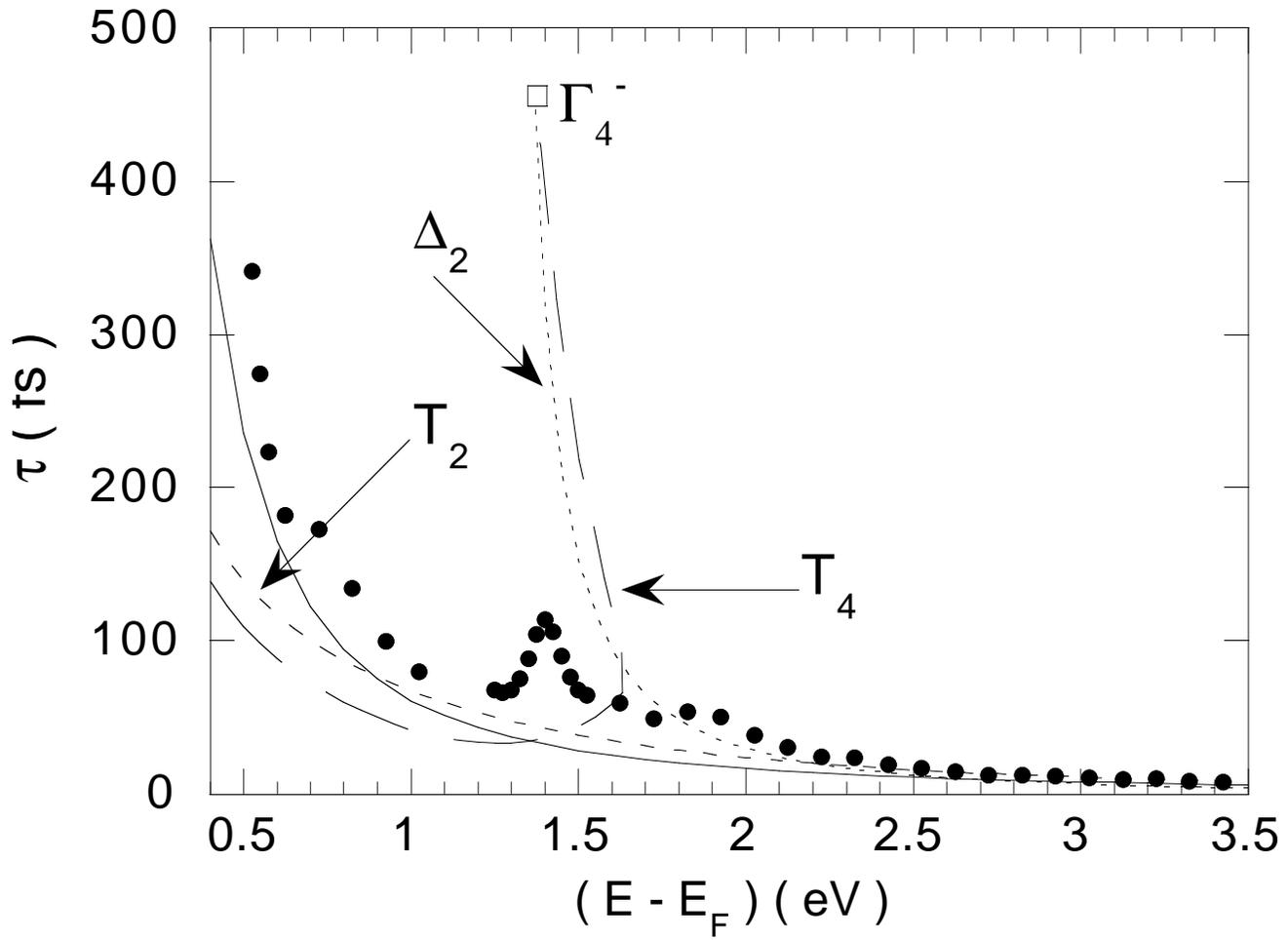
\begin{figure}
\caption[] {Hot-electron lifetimes in Be. Solid circles and the solid line
represent the same quantities as in Fig. 6. Short-dashed and long-dashed lines
represent the lifetime of hot electrons in bands $T_2$ and
$T_4$, respectively, with the wave vector along the $\Gamma K$ direction. The
dotted line represents the lifetime of hot electrons in the $\Delta_2$ band
along the
$\Gamma A$ direction. The open square is located at the energy of the level 
$\Gamma^{-}_4$, showing the beginning and the end of the 
$\Delta_2$ and ${\rm T}_4$ bands, respectively.}
\end{figure}

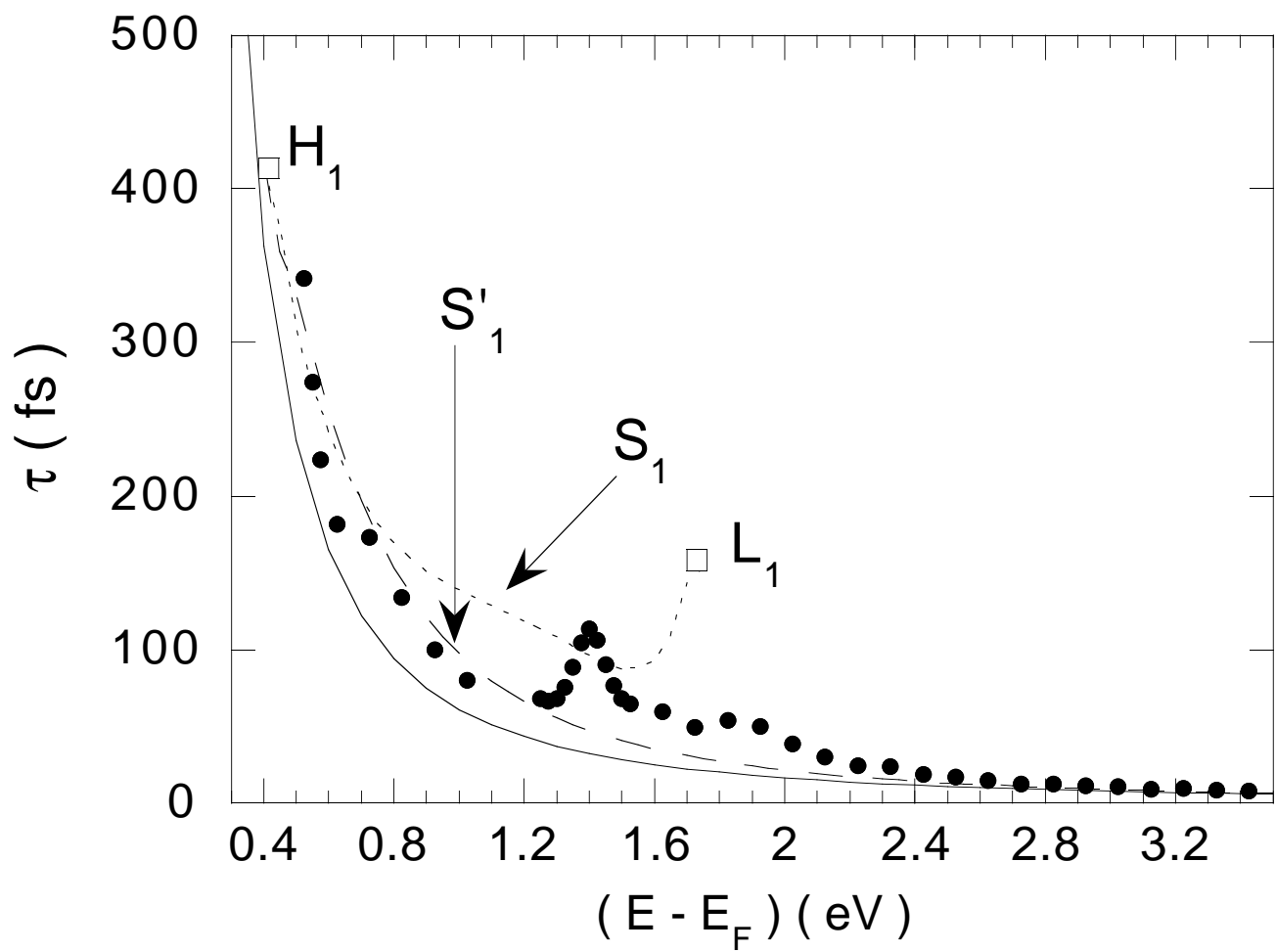
\begin{figure}
\caption[] {Hot-electron lifetimes in Be. Solid circles and the solid line
represent the same quantities as in Fig. 6. Dotted and dashed lines
represent the lifetime of hot electrons in bands $S_1$ and $S_1'$, respectively,
with the wave vector along the $LH$ and $HA$ directions. Open squares are located
at the energy of the levels $L_1$ and $H_1$.}
\end{figure}

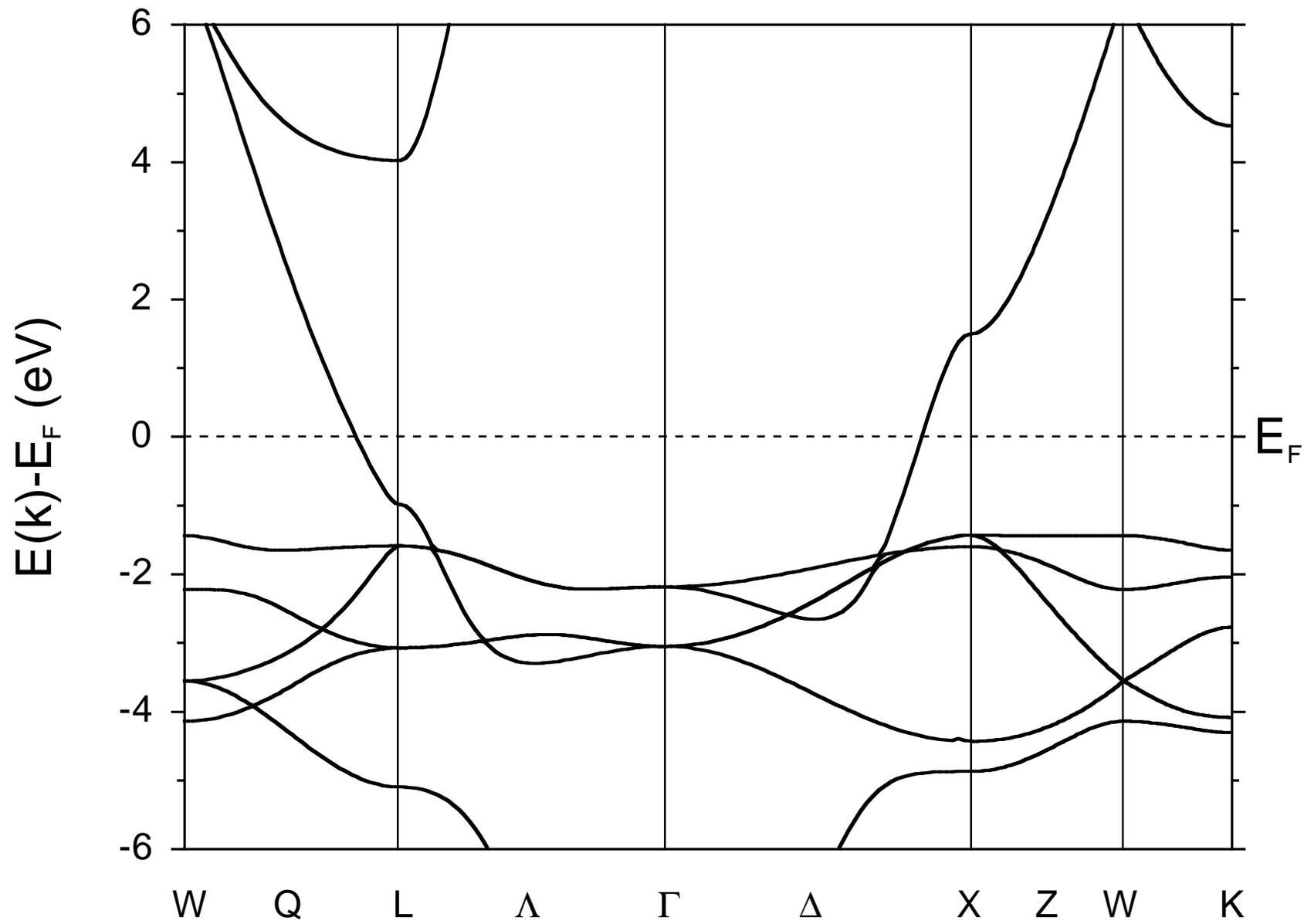
\begin{figure}
\caption[]{Calculated band structure of Cu along certain symmetry directions.}
\end{figure}

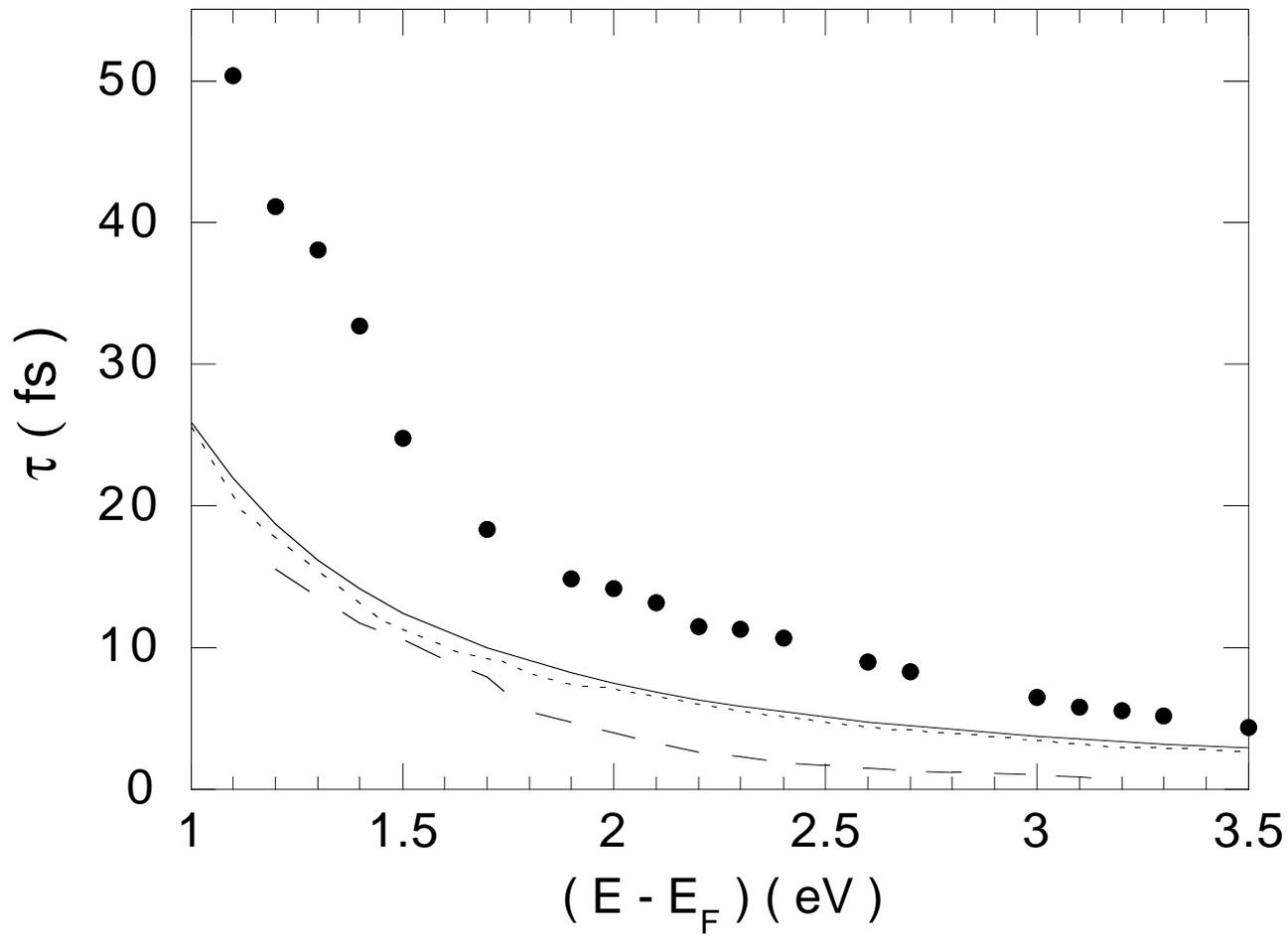
\begin{figure}
\caption[]
{Hot-electron lifetimes in Cu. As in Fig. 2 with $r_s=2.67$. The dashed line
represents the result of replacing all one-electron Bloch states by plane waves
(FEG calculation) but keeping in Eq. (\ref{eq18}) the actual number of states
available for real transitions.\cite{note2} The dotted line represents our full
{\it ab initio} calculation of $\tau(E)$, as obtained after averaging $\tau({\bf
k},n_i)$ of Eq. (\ref{eq10}) over wave vectors and over the band  structure for
each ${\bf k}$, but with the $3d$ shell assigned to the core in the
pseudopotential generation.}
\end{figure}


\begin{references}

\bibitem{QF} J. J. Quinn and R. A. Ferrell, Phys. Rev. {\bf 112}, 812
(1958).
\bibitem{Ritchie59} R. H. Ritchie, Phys. Rev. {\bf 114}, 644 (1959).
\bibitem{Quinn62} J. J. Quinn, Phys. Rev. {\bf 126}, 1453 (1962).
\bibitem{Ashley1} R. H. Ritchie and J. C. Ashley, J. Phys. Chem. Solids {\bf 26},
1689 (1965).
\bibitem{Kleinman} L. Kleinman, Phys. Rev. B {\bf 3}, 2982 (1971).
\bibitem{Shelton} J. C. Shelton, Surf. Sci. {\bf 44}, 305 (1974).
\bibitem{Penn0} D. R. Penn, Phys. Rev. B {\bf 13}, 5248 (1976).
\bibitem{Tung2} C. J. Tung, J. C. Ashley,  and R. H. Ritchie, Surf. Sci.
{\bf 81}, 427 (1979).
\bibitem{Penn1} D. R. Penn, Phys. Rev. B {\bf 22}, 2677 (1980).
\bibitem{Penn2} D. R. Penn, Phys. Rev. B {\bf 35}, 482 (1987).
\bibitem{Schmu} C. A. Schmutenmaer, M. Aeschlimann, H. E. Elsayed-Ali, R.
J. D. Miller, D. A. Mantell, J. Cao, and Y. Gao, Phys. Rev. B {\bf 50},
8957 (1994).
\bibitem{Hertel} T. Hertel, E. Knoesel, M. Wolf, and G. Ertl, Phys. Rev.
Lett. {\bf 76}, 535 (1996).
\bibitem{Petek0} S. Ogawa, H. Nagano and H. Petek, Phys. Rev. B {\bf 55},
1 (1997).
\bibitem{Cao1} J. Cao, Y. Gao, R. J. D. Miller, H. E. Elsayed-Ali, D. A.
Mantell, Phys. Rev. B {\bf 56}, 1099 (1997).
\bibitem{Aes1} M. Aeschlimann, M. Bauer, S. Pawlik, W. Weber, R.
Burgermeister, D. Oberli, and H. C. Siegmann, Phys. Rev. Lett. {\bf 79}, 5158
(1997).
\bibitem{Knoesel1} E. Knoesel, A. Hotzel and M. Wolf, Phys. Rev. B {\bf 57},
12812 (1998).
\bibitem{Goldm} A. Goldmann, R. Matzdorf, and F. Theilmann, Surf. Sci. {\bf 414}
, L932 (1998).
\bibitem{Cao2} J. Cao, Y. Gao, H. E. Elsayed-Ali, R. J. D. Miller, and D.
A. Mantell, Phys. Rev. B {\bf 58}, 10948 (1998).
\bibitem{Aes98} M. Bauer, S. Pawlik, M. Aeschlimann, Proc. of the SPIE 3272, 201
(1998).
\bibitem{Petek3} H. Petek and S. Ogawa, Prog. Surf. Sci. {\bf 56}, 239 (1998).
\bibitem{Igorprl} I. Campillo, J. M. Pitarke, A. Rubio, E. Zarate, and
P. M. Echenique, Phys. Rev. Lett. {\bf 83}, 2230 (1999).
\bibitem{Ekardt} W.-D. Sch\"one, R. Keyling, M. Bandi\'c, and W. Ekardt, Phys.
Rev. B {\bf 60}, 8616 (1999).
\bibitem{Kohn} P. Hohenberg and W. Kohn, Phys. Rev. {\bf 136}, B864
(1964); W. Kohn and L. Sham, Phys. Rev. {\bf 140}, A1133 (1965).
\bibitem{Fetter} A. L. Fetter and J. D. Walecka, {\it Quantum Theory of
Many-Particle Systems} (McGraw-Hill, New York, 1971).
\bibitem{Hedin} L. Hedin and S. Lundqvist, Solid State Phys. {\bf 23}, 1 (1969).
\bibitem{Runge} E. Runge and E. K. U. Gross, Phys. Rev. Lett. {\bf 52},
997 (1984).
\bibitem{Gross3} M. Petersilka, U. J. Gossmann and E. K. U. Gross,
Phys. Rev. Lett. {\bf 76}, 1212 (1996).
\bibitem{rs} The so-called electron-density parameter $r_s$ is defined by
the relation $1/n_0=(4/3)\pi r_s^3$, $n_0$ being the average electron density.
\bibitem{review} P. M. Echenique, J. M. Pitarke, E. V. Chulkov,
and A. Rubio, Chem. Phys. {\bf 251}, 1 (2000).
\bibitem{Silkin84} V. M. Silkin, E. V. Chulkov, I. Yu. Sklyadneva, 
and V. E. Panin, Sov. Phys. J. {\bf 27}, N 9, 762 (1984) [Izv. Vuzov. 
Fiz. {\bf 9} 56 (1984)]; E. V. Chulkov, V. M. Silkin, and E. N. 
Shirykalov, Phys. Met. and Metallogr. {\bf 64}, 1 (1987) [Fiz. Met. 
Metalloved. {\bf 64}, 213 (1987)].
\bibitem{Troullier} N. Troullier and J. L. Martins, Phys. Rev. B {\bf
43}, 1993 (1991).
\bibitem{Perdew} J. P. Perdew and A. Zunger, Phys. Rev.
B {\bf 23}, 5048 (1981).
\bibitem{Ceperley} D. M. Ceperley and B. J. Alder, Phys. Rev. Lett.
{\bf 45}, 1196 (1980).
\bibitem{LAPW} D. J. Singh, {\rm Plane Waves, Pseudopotentials, and the LAPW
Method} (Kluwer, Boston, 1994).
\bibitem{Igorcu} I. Campillo, A. Rubio, and J. M. Pitarke, Phys. Rev. B {\bf
59}, 12188 (1999).
\bibitem{Monk}  H. J. Monkhorst and J. D. Pack, Phys. Rev. B {\bf 13},
5188 (1976).
\bibitem{Platzman} P. M. Platzman, E. D. Isaacs, H. Williams, P.
Zschack, and G. E. Ice, Phys. Rev. B {\bf 46}, 12943 (1992).
\bibitem{Platzman2} W. Schulke, H. Schulte-Schrepping, and J. R. Schmitz,
Phys. Rev. B {\bf 47}, 12426 (1993).
\bibitem{Fleszar} A. Fleszar, A. A. Quong, and A. G. Eguiluz, Phys. Rev. Lett.
{\bf 74}, 590 (1995).
\bibitem{Fleszar2} B. C. Larson, J. Z. Tischler, E. D. Isaacs, P. Zschack, A. Fleszar, and A. G.
Eguiluz, Phys. Rev. Lett. {\bf 77}, 1346 (1997).
\bibitem{Igorstop} I. Campillo, J. M. Pitarke, and A. G. Eguiluz, Phys. Rev. B
{\bf 58}, 10307 (1998).
\bibitem{note1} The prediction of Eq. (\ref{eqQF}) (dashed line in the inset
of Figs. 2, 4, 6, and 10) would coincide, as $E\to E_F$, with the full
RPA free-electron gas calculation only in the high-density ($r_s\to 0$) limit.
\bibitem{Singhal} S. P. Singhal and J. Callaway, Phys. Rev. B {\bf 16}, 1744
(1977).
\bibitem{Segall2} F. Szumulowicz and B. Segall, Phys. Rev. B {\bf 21}, 5628
(1980).
\bibitem{Papa} D. A. Papaconstantopoulos, {\it Handbook of the band structure of
elemental solids} (Plenum Press, New York, 1986).
\bibitem{note2} We have followed Ogawa {\it et al}\,\cite{Petek0} to take the
electron-density parameter $r_s$ in the evaluation of $|\epsilon_{{\bf
G},{\bf G}}({\bf q},\omega)|^{-2}$ from the actual DOS at the Fermi level.
\bibitem{Adler} S. L. Adler, Phys. Rev. {\bf 130}, 1654 (1963).  

\end{references}
\end{document}